\documentclass{jfm}

\newcommand{\kk}{{\bm k}}
\newcommand{\pp}{{\bm p}}

\newcommand{\rr}{{\bm r}}

\newcommand{\uu}{{\bm u}}
\newcommand{\UU}{{\bm U}}

\usepackage{latexsym}
\usepackage{bm}
\usepackage{epstopdf, epsfig}
\usepackage{comment}
\usepackage{siunitx}

\usepackage[english]{babel}
\usepackage[T1]{fontenc}
\usepackage{endnotes}
\usepackage{amssymb,amsmath,amsfonts,amsbsy}
\usepackage{latexsym}
\usepackage{textcomp}
\usepackage{graphics,color}
\usepackage[colorlinks,citecolor=blue,urlcolor=blue,bookmarks=false,hypertexnames=true]{hyperref} 
\usepackage{graphicx}

\usepackage{float}
\usepackage[export]{adjustbox}
\usepackage[abs]{overpic}
\usepackage{xcolor,varwidth}

\usepackage{enumitem}

\usepackage{DejaVuSans}

\shorttitle{Interplay between Brownian and active tracer diffusion}
\shortauthor{H.~Nordanger, A.~Morozov, J.~Stenhammar}

\title{Interplay between Brownian and hydrodynamic tracer diffusion in suspensions of swimming microorganisms}

\author{Henrik Nordanger\aff{1},
	Alexander Morozov\aff{2},
	\and Joakim Stenhammar\aff{1}\corresp{\email{joakim.stenhammar@fkem1.lu.se}}}

\affiliation{\aff{1}Division of Physical Chemistry, Lund University, Box 124, S-221 00 Lund, Sweden
	\aff{2}SUPA, School of Physics and Astronomy, The University of Edinburgh, James Clerk Maxwell Building, Peter Guthrie Tait Road, Edinburgh, EH9 3FD, United Kingdom}

\begin{document}
	
\maketitle

\begin{abstract}
\noindent The general problem of tracer diffusion in non-equilibrium baths is important in a wide range of systems, from the cellular level to geographical lengthscales. In this paper, we revisit the archetypical example of such a system: a collection of small passive particles immersed in a dilute suspension of non-interacting dipolar microswimmers, representing bacteria or algae. In particular, we consider the interplay between thermal (Brownian) diffusion and hydrodynamic (active) diffusion due to the persistent advection of tracers by microswimmer flow fields. Previously, it has been argued that even a moderate amount of Brownian diffusion is sufficient to significantly reduce the persistence time of tracer advection, leading to a significantly reduced value of the effective active diffusion coefficient $D_A$ compared to the non-Brownian case. Here, we show by large-scale simulations and kinetic theory that this effect is in fact only practically relevant for microswimmers that effectively remain stationary while still stirring up the surrounding fluid, so-called \emph{shakers}. In contrast, for moderate and high values of the swimming speed $v_s$, relevant for biological microswimmer suspensions, the effect of Brownian motion on $D_A$ is negligible, leading to the effects of advection by microswimmers and Brownian motion being additive. This conclusion contrasts with previous results from the literature, and encourages a reinterpretation of recent experimental measurements of $D_A$ for tracer particles of varying size in bacterial suspensions. 
\end{abstract}

\section{Introduction} \label{section:introduction}
\begin{figure*}
\centering

\begin{minipage}[t]{0.4\textwidth}
\includegraphics[width=\textwidth, clip, trim=0.3cm 0.3cm 0.3cm 0.3cm]{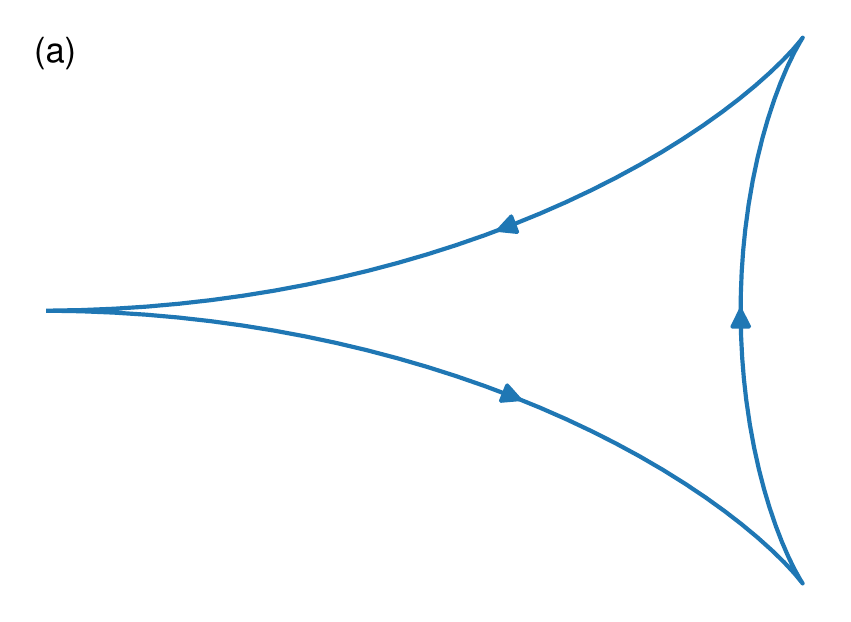}
\end{minipage}
\begin{minipage}[t]{0.4\textwidth}
    \includegraphics[width=\textwidth, clip, trim=0.3cm 0.3cm 0.3cm 0.3cm]{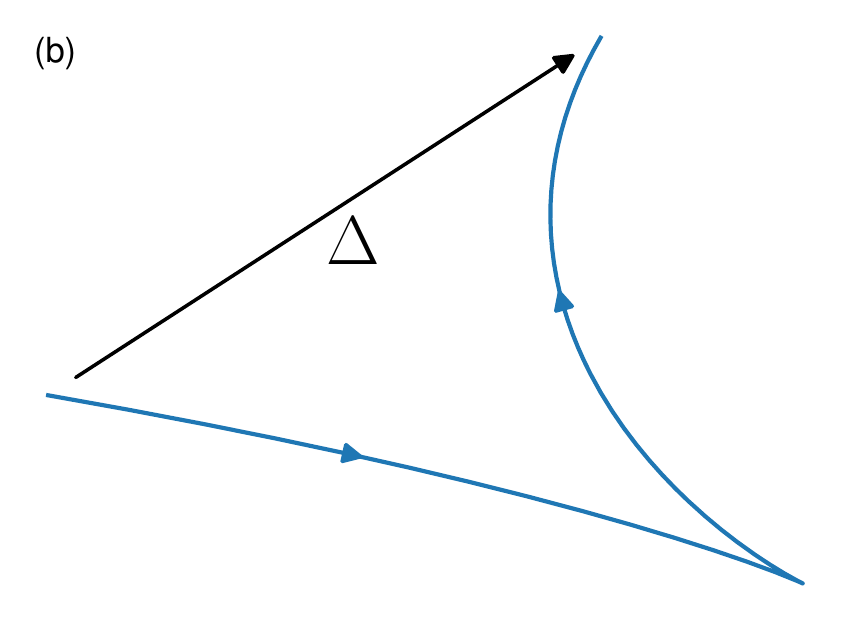}
\end{minipage}

\caption{\textbf{Tracer trajectories for infinite and finite swimmer paths.} Panel (a) shows a typical trajectory for a non-diffusing tracer advected by a non-tumbling, point-dipole swimmer following an effectively infinite, straight path, while (b) shows the corresponding trajectory terminated due to a tumbling event. Note that, per Eq.~\eqref{eq:DA_approx}, the effective tracer diffusion $D_A$ is independent of tumbling rate $\lambda$ for sufficiently high swimming speeds when averaged over all possible swimmer-tracer configurations even though the net displacement $\Delta$ is much larger for the tumbling swimmer. Tracer trajectories were obtained through direct numerical integration using a non-regularised dipolar flow field, as described by~\cite{Morozov1}.} 
\label{fig:trajectories}
\end{figure*}

\noindent Understanding the mass transport of colloidal and molecular species in non-equilibrium environments is crucial for various processes, ranging from active intracellular transport~\citep{Koslover:PhysBiol:2020} to the dispersion of nutrients in world oceans~\citep{Katija:ExpBiol:2012}. Apart from its practical importance, the transport properties of tracer particles in generic ``active baths'' has attracted much interest from a statistical physics perspective, where they can be viewed as a minimal example of particles driven by external, non-equilibrium noise~\citep{Volpe:PRE:2016,Park:SoftMatter:2020}. Beyond the level of tracer particles driven by generic non-equilibrium noise, the archetypical example of a tracer particle in an active bath is a collection of point-like tracers being advected by a set of microswimmers such as bacteria or algae~\citep{Lauga1}. When swimming through a viscous fluid, these swimmers create long-ranged flow fields that advect the tracers, leading to tracer dynamics that is ballistic at short times and diffusive over timescales longer than the autocorrelation time of the local flow field~\citep{Lin1}. Realisations of this system have been extensively studied both experimentally, typically in suspensions of \emph{E. coli} bacteria~\citep{Wu_Libchaber:PRL:2000,Drescher1,Jepson1,Kim1,Koumakis1,Mino1,Mino2,Patteson1,Peng1,Semeraro1}  or \emph{Chlamydomonas} algae~\citep{Leptos1,Ortlieb1,Yang1,Eremin:2021}, and theoretically, with microswimmers typically being modelled either as force dipoles acting on the surrounding fluid~\citep{Morozov1,Yeomans:JFM:2013,Pushkin1,Nordanger:PRFluids:2022}, as spherical ``squirmers'' with an imposed slip velocity along their body~\citep{Lin1,Thiffeault2,Thiffeault1}, or as needle-shaped ``slender swimmers'' with imposed stresses along their body lengths~\citep{Krishnamurthy1,Saintillan2}. While the details of these three microswimmer models differ, the results regarding enhanced tracer diffusion are largely generic and consistent with experimental results, which have shown the swimmer-induced, hydrodynamic diffusivity $D_A$ to scale linearly with microswimmer density $n$ in the dilute limit where swimmer-swimmer correlations can be neglected~\citep{Lin1,Thiffeault2,Mino1}. In this limit, a fruitful way of calculating $D_A$ is to consider the net displacement due to binary swimmer-tracer scattering events~\citep{Pushkin1,Morozov1}; two examples of resulting (deterministic) tracer trajectories for scattering events are shown in Fig.~\ref{fig:trajectories}. For a non-tumbling swimmer, starting and ending at $x = \pm \infty$, the resulting loop is closed, leading to a vanishing tracer net displacement $\Delta$ (Fig.~\ref{fig:trajectories}a). For tumbling swimmers with a finite persistence length, the trajectory is however punctuated mid-way through the tracer loop, leading to significantly larger values of $\Delta$ (Fig.~\ref{fig:trajectories}b). The resulting value of $D_A$ due to a large set of such scattering events can then be obtained by explicitly summing over all possible sets of scattering parameters. 

In spite of the dependence of $\Delta$ on the microswimmer tumbling rate $\lambda$ for single scattering events such as that in Fig.~\ref{fig:trajectories}, \cite{Pushkin1} showed that, in the limit of large swimming speeds $v_s$, $D_A$ is in fact \emph{independent} of $\lambda$ when summed over all possible swimmer trajectories. This result was later generalised by \cite{Morozov:PRX:2020} to arbitrary swimming speeds, leading to the following approximate expression for $D_A$:
\begin{equation}\label{eq:DA_approx}
    D_A \approx \frac{7\kappa^2 n}{2048 \lambda \varepsilon + 336\pi v_s}.
\end{equation}
Here, $\kappa$ is the magnitude of the microswimmer dipole (in units of volume over time) and $\varepsilon$ is a characteristic linear size of the microswimmer, which we take to be equal to the short-range regularisation length of the dipolar flow field; unless stated otherwise, we will in the following use $\kappa$ and $\varepsilon$ to non-dimensionalise the numerical data. 

Rather than using the scattering approach outlined above, Eq.~\eqref{eq:DA_approx} was derived by formulating a kinetic theory for the spatiotemporal correlations of the disturbance velocity field $\UU$ created by the swimmers. Due to the linearity of Stokes flow, we can write $\UU$ as the superposition of the individual swimmer flow fields $\uu_s$:
\begin{equation}\label{eq:Pairwise_U}
\UU(\rr,t) = \sum_{i=1}^{N} \uu_s(\rr;\rr_i,\pp_i),   
\end{equation}
where $\rr_i$ and $\pp_i$ is, respectively, the position and orientation of swimmer $i$. Knowing the statistical properties of $\UU$, $D_A$ can readily be calculated \emph{via} the Green-Kubo relation
\begin{align}\label{eq:GreenKubo}
D_A = \frac{1}{3} \int_{0}^{\infty}\langle \dot{\rr}_T(t) \cdot \dot{\rr}_T(0)\rangle dt =  \frac{1}{3} \int_{0}^{\infty}\langle \UU(\rr_T,t) \cdot \UU(\rr_T,0) \rangle dt \equiv \frac{1}{3} \int_{0}^{\infty}C_T(t) dt,
\end{align}
where, in the second equality, we have assumed point-like tracers advected by the disturbance flow, so that $\dot{\rr}_T = \UU(\rr_T)$, and the third equality defines the velocity autocorrelation function $C_T(t)$ in the co-moving tracer frame. While yielding identical results for dilute suspensions as the scattering approach discussed above, kinetic theories are however more readily extended to accommodate the effect of swimmer-swimmer correlations due to the mutual advection and reorientation of swimmers~\citep{Morozov:PRX:2020}. Importantly, these interactions break the symmetry between rear-actuated (pusher) swimmers, such as most bacteria, and front-actuated (puller) ones, such as \emph{Chlamydomonas}, leading to a super-linear scaling of $D_A$ with $n$ for pushers and a corresponding sub-linear scaling for pullers~\citep{Stenhammar1}. 

Equation~\eqref{eq:DA_approx} shows two qualitatively different regimes for high and low $v_s$: For $v_s  \rightarrow 0$ -- the so-called  \emph{shaker} limit -- the dominant mechanism controlling the decorrelation of $C_T(t)$ is tumbling of the swimmer. For fast swimmers, with $v_s \gg \lambda \varepsilon$, the decorrelation of the fluid velocity is instead dominated by the swimmer self-propulsion and thus independent of $\lambda$, and Eq.~\eqref{eq:DA_approx} reduces to the expression derived by~\cite{Pushkin1}. A third, somewhat less explored, mechanism affecting $C_T(t)$ is Brownian translational diffusion of the tracer: even though Brownian diffusion does not affect the statistics of the flow field $\UU(\rr,t)$ as measured in the lab frame, the positional noise of the tracer particle will cause it to cross the streamlines of the disturbance flow, thus perturbing its trajectory compared to the athermal case shown in Fig. \ref{fig:trajectories} and leading to a lower $D_A$. Since the Brownian diffusion constant $D_0$ depends inversely on the tracer radius per the Stokes-Einstein relation, the magnitude of this effect is expected to be significant primarily for small tracer particles, and it has been hypothesised to explain the non-monotonic size dependence of enhanced tracer diffusion seen in experiments with colloids in \emph{E. coli} suspensions, where \cite{Patteson1} observed a maximum in $D_A$ for a tracer radius of approximately 5 $\mu$m. The effect of Brownian motion on enhanced tracer diffusion was furthermore theoretically analysed by \cite{Kasyap2014} for a model of slender-body swimmers, showing that $D_A$ is a non-monotonic function of $D_0$, with $D_A$ first showing a small increase for intermediate $D_0$, before falling below the athermal value as $D_0$ grows larger. In a more recent study of finite-size spherical tracers in microswimmer suspensions, \cite{Dyer:PoF:2021} numerically analysed the combined effect of thermal fluctuations and near-field flows on the size-dependent tracer dynamics, finding a similar non-monotonic behaviour as observed experimentally. In this Paper, we will revisit the problem of the interplay between Brownian and hydrodynamic diffusion for the simple case of point-like tracers immersed in a dilute suspension of microswimmers described via a regularised dipolar flow field. Using kinetic theory and large-scale lattice Boltzmann simulations of \emph{E. coli}-like suspensions, we show that the effect of Brownian diffusion on active diffusion is only practically relevant whenever $v_s < \lambda \varepsilon$, which corresponds to extremely slow (or frequently tumbling) swimmers. For biologically relevant values of $v_s$ and $\lambda$, swimming is instead the dominant decorrelation mechanism, so that $D_A$ becomes independent of both $D_0$ and $\lambda$. In contrast to previous studies, our results thus indicate that the effect of Brownian motion on the enhanced diffusion is in fact negligible for most microswimmer realisations, and thus that the experimentally observed non-monotonic size dependence of $D_A$ on tracer size reported by~\cite{Patteson1} has other explanations. 

\section{Model and Method}\label{section:method}
\noindent We consider a collection of $N$ non-interacting microswimmers at number density $n = N/V$ moving through a three-dimensional viscous fluid of viscosity $\mu$. Each microswimmer is composed of two equal and opposite point forces of magnitude $F$ separated by a length $\ell$ and swims with a constant speed $v_s$. The resulting reduced hydrodynamic dipole strength is $\kappa = F\ell/\mu$. The swimming direction $\pp_i$ furthermore relaxes through Poisson-distributed random tumbles with uncorrelated directions occurring with average frequency $\lambda$. 

The position $\rr_T$ of a point-like tracer obeys the equation of motion
\begin{equation}\label{eq:rdot_tr}
    \dot{\rr}_T = \UU(\rr_T) + \sqrt{2D_0}\boldsymbol{\eta},
\end{equation}
where $\boldsymbol{\eta}$ is a unit-variance white noise, $\delta$-correlated in time, and $D_0$ is the Brownian diffusion constant. Thus, the effect of Brownian motion is fully contained in the tracer dynamics, while we assume the effects of thermal fluctuations on the fluid and on the pairwise swimmer-tracer dynamics to be subdominant. The fluid disturbance velocity $\UU(\rr_T)$ due to the presence of all swimmers is given by Eq.~\eqref{eq:Pairwise_U} and can, in principle, be explicitly summed up on each timestep. However, to avoid the (prohibitively costly) pairwise summation over all swimmers and tracers, we instead numerically solve for the flow field using an efficient point-force implementation of the lattice Boltzmann (LB) method described previously~\citep{Bardfalvy:PRL:2020,Nash1}. Our simulations comprise a system with periodic boundaries and a size of $100^3$ lattice units. In LB units, set by the LB lattice spacing $\Delta l$ and time step $\Delta t$, the microswimmer density was kept constant at $n = 0.01$ unless otherwise stated, corresponding to $N = 10^4$ microswimmers. Furthermore, $N_T = 5 \times 10^4$ point tracers were included for statistical averaging. In addition to the LB simulations, in Section~\ref{sec:Theory} we will furthermore extend the kinetic theory developed previously by~\cite{Morozov:PRX:2020} to the case of microswimmers undergoing Brownian diffusion and show that, in the limit of non-interacting swimmers, the derived expression for $D_A$ is equivalent to that for Brownian tracers in a suspension of non-Brownian microswimmers.

Inserting Eq.~\eqref{eq:rdot_tr} into the Green-Kubo relation~\eqref{eq:GreenKubo} yields
\begin{equation}
    D_{\mathrm{Tot}} = D_0 + \frac{1}{3} \int_{0}^{\infty} C_T(t) dt = D_0 + D_A, 
\end{equation}
where $C_T$ was defined in Eq.~\eqref{eq:GreenKubo}. Thus, to obtain $D_A$, we numerically evaluate the time correlation of the disturbance velocity measured in the co-moving tracer frame. Since the tracer position $\rr_T$ depends on $D_0$, $C_T(t)$, and thus $D_A$, will depend implicitly on $D_0$. 

To characterise the system, we will use three dimensionless quantities. Firstly, we define the P\'eclet number, which measures the relative importance of active and thermal forces, as
\begin{equation}\label{eq:Pe_Def}
    \mathrm{Pe} \equiv \frac{D_A(D_0=0)}{D_0},
\end{equation}
where $D_A(D_0=0)$ is the active diffusivity of a tracer immersed in an equivalent microswimmer suspension but in the absence of Brownian tracer motion. It should be noted that our definition of Pe is qualitatively different from that of \cite{Kasyap2014}, who, instead of $D_A(D_0=0)$, use the swimming speed $v_s$ to characterise the active forces. We however argue that the activity experienced by the tracers depend on the magnitude of the velocity fields generated by the swimmers, and is thus dependent on $\kappa$ and encoded in $D_A(D_0 = 0)$, while $v_s$ is instead a measure of the swimmers' self-propulsion. For experimental realisations of microswimmers, $\kappa$ and $v_s$ are directly proportional to each other; however, the specific relation between between them will nevertheless be specific to each type (or species) of swimmer, and decoupling them conveniently enables us to study separately the effects of self-propulsion and fluid forcing, as we demonstrate further below. 

Secondly, we measure the change in active diffusion due to Brownian motion through the quantity
\begin{equation}
    \xi \equiv \frac{D_A(D_0)}{D_A(D_0 = 0)}. 
\end{equation}
In the limit $\mathrm{Pe} \rightarrow \infty$, where Brownian motion becomes negligible, we thus expect that $\xi \rightarrow 1$ as $D_A$ approaches its non-Brownian value. Finally, in accordance with \cite{Morozov:PRX:2020}, we account for the effect of microswimmer self-propulsion using the reduced swimmer persistence length $L$, defined by
\begin{equation}\label{eq:L_def}
    L \equiv \frac{v_s}{\varepsilon \lambda}.
\end{equation}
 
\section{Kinetic theory}\label{sec:Theory}

\noindent In this Section, we will outline the main steps in the derivation of $D_A$ for a suspension of Brownian tracer particles immersed in a dilute microswimmer suspension, whose dynamics are governed by Eq.~\eqref{eq:rdot_tr}. Just as in our previous works~\citep{Morozov:PRX:2020,Stenhammar1} , we describe the flow field measured at $\rr$ due to a swimmer with position $\rr_i$ and orientation $\pp_i$ by a regularised dipolar flow field $\uu_s(\rr)$
\begin{equation}\label{eq:u_s}
    \uu_s(\rr;\rr_i,\pp_i) = \frac{\kappa}{8\pi} \left[ 3\frac{(\pp_i \cdot \rr')^2 \rr' + \varepsilon^2 (\pp_i \cdot \rr')\pp_i }{(r'^2 + \varepsilon^2)^{5/2}} - \frac{\rr'}{(r'^2 + \varepsilon^2)^{3/2}} \right],
\end{equation}
where $\rr' = \rr-\rr_i$, $r' = |\rr'|$, and $\varepsilon$ is the regularisation length. Our starting point is the derivation of \cite{Morozov:PRX:2020}, where we formulated and solved a kinetic theory describing the fluctuations of the velocity field $\UU(\rr,t)$ due to a superposition of single-swimmer flow fields. In the limit of non-interacting swimmers, which is the case that we consider here, the temporal correlations of the steady-state velocity field $\UU$ measured in the lab frame, $C_U(t) \equiv \langle \UU(\rr,t) \cdot \UU(\rr,0) \rangle$, is given by 
\begin{equation}\label{eq:CU_nonBrownian}
    C_U(t) = \frac{\kappa^2 n}{15 \pi^2 \varepsilon} \int_0^{\infty} A^2(\zeta) e^{-\tau} \mathcal{F}(L\zeta\tau) d\zeta,
\end{equation}
where
\begin{equation}
    A(x) = \frac{1}{2} x^2 K_2(x),
\end{equation}
with $K_2$ being the modified Bessel function of the second kind, and
\begin{equation}
    \mathcal{F}(x) = 15\frac{(5x^2-12)\sin x - x(x^2-12)\cos x}{x^5},
\end{equation}
defined such that $\mathcal{F}(0) = 1$. We furthermore used the dimensionless variables $L = v_s/(\lambda \varepsilon)$, $\tau = t\lambda$, and $\zeta = k\varepsilon$, where $k = |\kk|$ is the  wavevector magnitude. Equation~\eqref{eq:CU_nonBrownian} can equivalently be expressed in closed form in terms of elliptic integrals, as given by Eq.~(72) of \cite{Morozov:PRX:2020}. Equation~\eqref{eq:CU_nonBrownian} contains decorrelation of the flow field due to two separate mechanisms: exponential decay of $C_U(t)$ due to tumbling, and a more complex, oscillatory behaviour due to swimming, encoded in the function $\mathcal{F}$. 

To obtain the hydrodynamic diffusivity of a passive tracer, \cite{Morozov:PRX:2020} used a \emph{stationary tracer} approximation, implying that the tracer advection by the swimmer flow field is negligible compared to the self-propulsion of the swimmer. This implies that $\rr_T$ remains effectively constant over the time it takes for $\UU$ to relax, such that $C_T(t) = \langle \UU(\rr_T[t],t) \cdot \UU(\rr_T[t=0],0) \rangle \approx \langle \UU(\rr_T[t=0],t) \cdot \UU(\rr_T[t=0],0) \rangle = C_U(t)$,
where, in the last equality, we have made the additional assumption that the tracers are homogeneously distributed in space so that they sample an unweighted spatial average of the flow field. Thus, if we can replace the correlation function $C_T$ in the co-moving tracer frame with that in the stationary lab frame, $C_U$, we can insert Eq.~\eqref{eq:CU_nonBrownian} into the Green-Kubo relation~\eqref{eq:GreenKubo} and integrate over time to yield the following expression for $D_A$, identical to Eq. (85) of \cite{Morozov:PRX:2020}:
\begin{equation}\label{eq:DA_nonBrownian}
    D_A = \frac{\kappa^2 n}{45 \pi^2 \lambda \varepsilon} \int_0^{\infty} A^2(\zeta)  \mathcal{G} ( L\zeta ) d\zeta,
\end{equation}
where
\begin{equation}
    \mathcal{G}(x) = \frac{5}{2} \frac{3x + 2x^3 -3(1+x^2)\arctan x}{x^5},
\end{equation}
defined such that $\mathcal{G}(0) = 1$. By matching the asymptotic behaviours for $L\rightarrow 0$ and $L \rightarrow \infty$, Equation~\eqref{eq:DA_nonBrownian} can furthermore be approximated by the simple expression given in Eq.~\eqref{eq:DA_approx}

The generalisation of Eq.~\eqref{eq:DA_nonBrownian} to the case of Brownian tracers might seem straightforward, but unfortunately is not: Since the approximation $C_T(t) = C_U(t)$ amounts to the tracer remaining effectively stationary during a swimmer-tracer scattering event, this approximation will, by construction, not capture any effects on $D_A$ coming from Brownian diffusion across streamlines. This can easily be realised by noticing that $C_U(t)$ is solely a property of the swimmer suspension, and will be strictly unaffected by the tracer dynamics; thus, $D_A$ in Eq.~\eqref{eq:DA_nonBrownian} remains unaffected by the inclusion of tracer diffusion. Instead of going beyond the stationary tracer approximation, we circumvent this problem by noticing that, in a suspension of non-interacting microswimmers, the dynamics of a Brownian tracer will be \emph{statistically identical} to that of a non-Brownian tracer in a suspension of Brownian \emph{swimmers} with the same translational diffusivity $D_0$. This is because the single-swimmer flow field in Eq.~\eqref{eq:u_s} solely depends on the separation vector $\rr_T - \rr_i$, implying that the flow field experienced by a diffusing tracer (\emph{i.e.}, noise acting on $\rr_T$) is identical to that experienced by a non-Brownian tracer sampling the flow field from a swimmer with the same noise instead applied to $\rr_i$. In a non-interacting microswimmer suspension this equivalence is exact as long as the noise has zero mean and identical spectral properties, and we verify it numerically in Fig.~\ref{fig:xi_Pe}. It however breaks down as soon as swimmer-swimmer correlations become significant since swimmer diffusion will then affect the magnitude of such correlations, which tracer diffusion will not.

Thus, as outlined in Appendix \ref{App:CT}, we instead calculate $C_U(t)$ for the case of a suspension of diffusing swimmers, described by the dynamics
\begin{equation}
    \dot{\rr}_i = v_s \pp_i + \sqrt{2D_0}\boldsymbol{\eta},
\label{eq:eom}
\end{equation}
including the same tumbling dynamics as before. This yields the following generalisation of Eq.~\eqref{eq:CU_nonBrownian}:
\begin{equation}\label{eq:CU}
    C_U(t;D_0) = \frac{\kappa^2 n}{15 \pi^2 \varepsilon} \int_0^{\infty} A^2(\zeta) e^{-(1+\tilde{D}\zeta^2) \tau} \mathcal{F}(L\zeta\tau) d\zeta,
\end{equation}
where we have additionally defined the non-dimensional diffusivity $\tilde{D} = D_0/(\lambda \varepsilon^2)$. Since the effect of Brownian diffusion is now fully incorporated into the properties of $C_U$, we again use the stationary-tracer approximation and insert this expression into the Green-Kubo relation~\eqref{eq:GreenKubo}, leading to the following expression for $D_A$ in the presence of Brownian diffusion:
\begin{equation}\label{eq:DA_exact}
    D_A(D_0) = \frac{\kappa^2 n}{45 \pi^2 \lambda \varepsilon} \int_0^{\infty} \frac{A^2(\zeta)}{1 + \tilde{D}\zeta^2}  \mathcal{G}\left( \frac{L\zeta}{1 + \tilde{D}\zeta^2} \right) d\zeta.
\end{equation}
In Section~\ref{sec:Results}, we numerically evaluate Eqs.~\eqref{eq:CU} and~\eqref{eq:DA_exact} and compare the results with direct numerical simulations of microswimmer suspensions. 

\section{Results and Discussion}\label{sec:Results}

\begin{figure*}
\centering

\begin{minipage}[t]{0.49\textwidth}
\includegraphics[width=\textwidth, clip, trim=0.0cm -0.1cm 0.0cm 0cm]{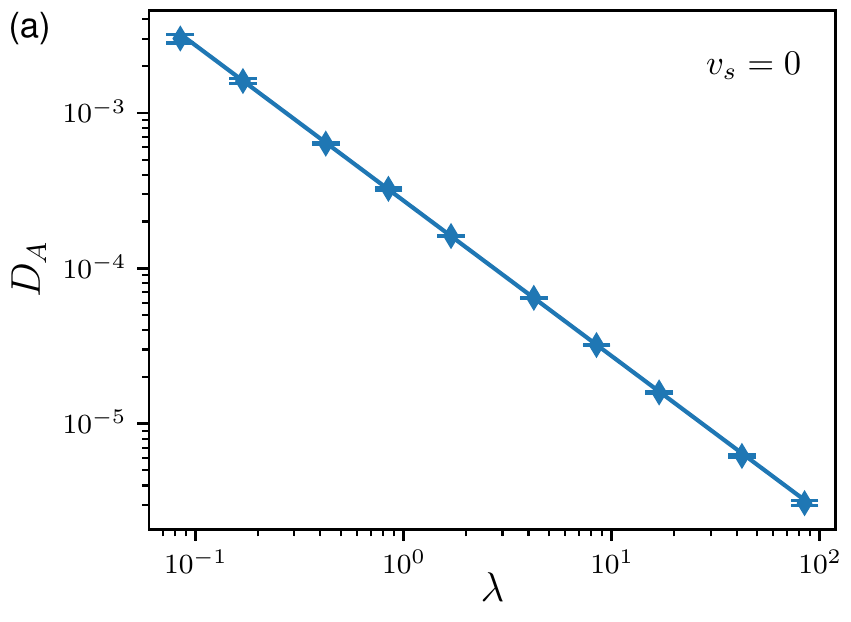}
\end{minipage}
\hfill
\begin{minipage}[t]{0.49\textwidth}
    \includegraphics[width=\textwidth, clip, trim=0.1cm 0cm -0.05cm 0cm]{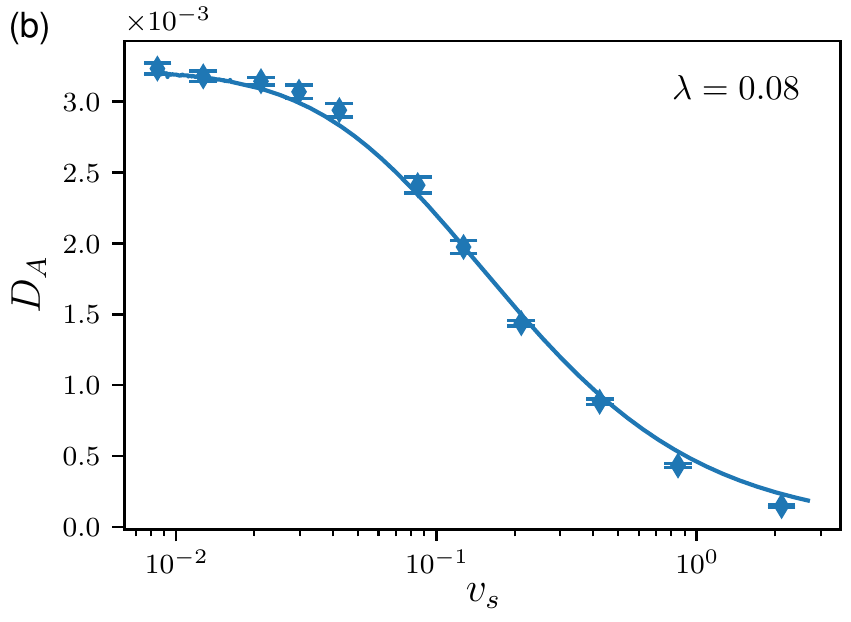}
\end{minipage}

\caption{\textbf{Effective tracer diffusion in the absence of Brownian motion}. Panel (a) shows the $\lambda^{-1}$-dependence of $D_A(D_0=0)$ for shakers with $v_s = 0$, and panel (b) its $v_s$-dependence at constant $\lambda = 10^{-4}$. Symbols denote simulation results and solid lines show results from Eq.~\eqref{eq:DA_nonBrownian} using $\varepsilon$ New{as a fitting parameter}. Error bars represent one standard deviation as obtained from averaging over four separate runs with different initial conditions. The results are nondimensionalised in terms of $\kappa$ and $\varepsilon$. } 
\label{fig:D_lambda_vs}
\end{figure*}

In Fig.~\ref{fig:D_lambda_vs} we begin by verifying Eq.~\eqref{eq:DA_nonBrownian} for the hydrodynamic diffusion coefficient $D_A$ in the limit $D_0 = 0$. In Fig.~\ref{fig:D_lambda_vs}a, we demonstrate the $\lambda^{-1}$ dependence of $D_A$ in the shaker limit $v_s = 0$, while Fig.~\ref{fig:D_lambda_vs}b shows its more complex dependence on $v_s$ for constant $\lambda$. Apart from numerically verifying the kinetic theory expression~\eqref{eq:DA_nonBrownian}, these results illustrate how $D_A$ decreases abruptly due to the temporal decorrelation of the flow field $\UU(\rr_T,t)$ induced respectively by tumbling and swimmer self-propulsion. The slight deviation between the theoretical curve and simulation results at small $v_s$ in Fig.~\ref{fig:D_lambda_vs}b is likely due to the specific form for the short-range regularisation, which becomes important as $v_s \rightarrow 0$. In our derivation of Eq.~\eqref{eq:DA_nonBrownian} we use the regularised flow field~\eqref{eq:u_s}, based on the regularisation first introduced by~\cite{Cortez2005}. In the LB simulations we instead use a numerical interpolation scheme based on a regularisation of the $\delta$ function~\citep{Peskin1} acting separately on the two point forces that make up each microswimmer. Unlike the expression in~\eqref{eq:u_s}, this numerical regularisation does not allow a direct mapping (or adjustment) of the regularisation length $\varepsilon$. We thus do not expect perfect agreement between kinetic theory and simulation in the low-$v_s$ regime where the short-range regularisation becomes important, and therefore treat $\varepsilon$ as a fitting parameter when comparing data from LB simulations with kinetic theory predictions. However,  we find that the fitted value of $\varepsilon$ only varies slightly ($\varepsilon \in [1.9 \Delta l, 2.5\Delta l]$) for the values of $v_s$ used throughout this work, in good accordance with the fact that the regularised $\delta$ function is interpolated over a support of $2 \Delta l$ in each Cartesian direction; for a more in-depth discussion of the effect of the interpolation scheme on the tracer dynamics, see~\cite{deGraaf1}. 

\begin{figure}
\centering

\begin{minipage}[t]{0.49\textwidth}
\includegraphics[width=\textwidth, clip, trim=0.0cm -0.1cm 0.0cm 0cm]{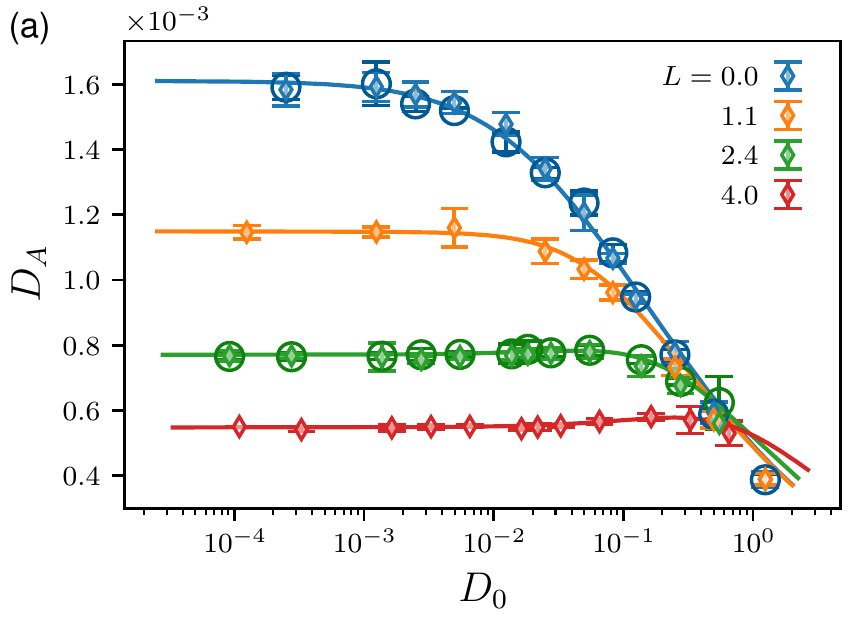}
\end{minipage}
\hfill
\begin{minipage}[t]{0.49\textwidth}
\includegraphics[width=\textwidth, clip, trim=0.0cm -0.1cm 0.0cm 0cm]{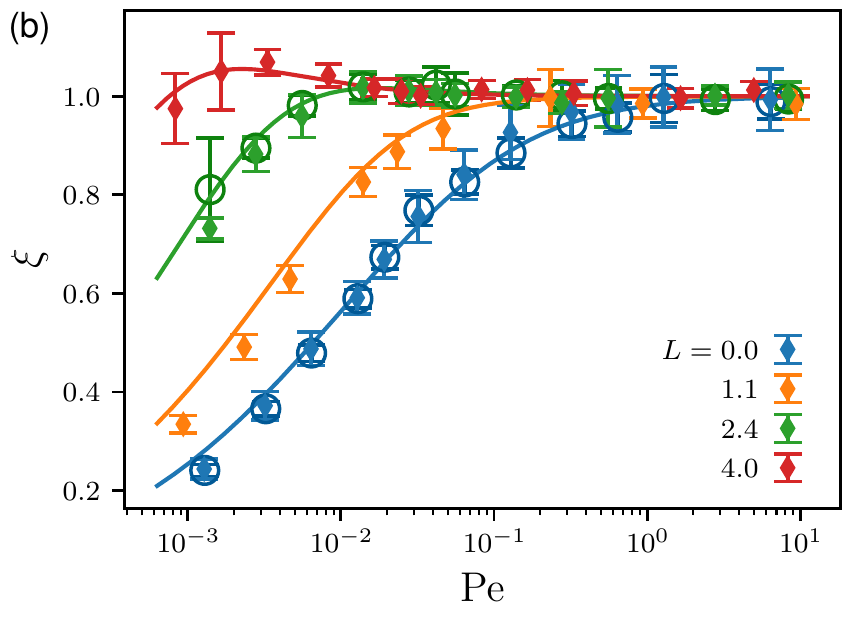}
\end{minipage}

\caption{\textbf{Brownian motion suppresses active diffusion for slow swimming speeds.} Panel (a) shows values of $D_A$ measured from LB simulations (diamonds) and calculated from Eq.~\eqref{eq:DA_exact} (solid lines), both expressed in LB units. Panel (b) shows the same data but expressed in the dimensionless quantities $\xi$ and Pe. For very slow swimmers with $L \lesssim 1$, $D_A$ is reduced compared to the non-Brownian value ($\xi = 1$) whenever $\mathrm{Pe} < 1$, while for faster swimmers, significantly lower values of Pe are necessary to affect $D_A$. The circles for $L = 0$ and 2.4 correspond to the hydrodynamic diffusion of non-Brownian tracers measured in a suspension of Brownian \emph{swimmers} of the same $D_0$, verifying the statistical equivalence between tracer and swimmer diffusion in the non-interacting limit. Error bars represent one standard deviation as obtained from averaging over four separate runs with different initial conditions. The results in panel (a) are nondimensionalised using $\kappa$ and $\varepsilon$. } 
\label{fig:xi_Pe}
\end{figure}

In Fig.~\ref{fig:xi_Pe}, we study the additional effect of varying the Brownian diffusion coefficient $D_0$, as encoded in Eq.~\eqref{eq:DA_exact}. From the data in Fig.~\ref{fig:xi_Pe}a, it is clear that, for large enough $D_0$, the active diffusivity $D_A$ decreases compared to its non-Brownian value. To enable an easier analysis of the effect of varying swimming speed, in panel (b) we present the same data instead plotted as a function of the reduced variables $\xi$ and Pe. For shakers with $L = 0$ (blue curve in Fig.~\ref{fig:xi_Pe}b), $D_A$ is reduced compared to its non-Brownian value ($\xi < 1$) as soon as $\mathrm{Pe} < 1$, reaching a value as low as $\xi = 0.2$ for $\mathrm{Pe} \approx 10^{-3}$. For finite values of $v_s$, this effect on $D_A$ however occurs for gradually lower values of Pe; for the fastest swimmers considered here, with $L = 4.0$, no significant reduction of $D_A$ is observed even for Pe  as low as $10^{-3}$. Instead, we observe a small but significant \emph{increase} in the active diffusion compared to its non-Brownian value, in accordance with what was previously observed for slender swimmers by~\cite{Kasyap2014}; we discuss this effect further below. Crucially, a reduced persistence length $L = 4$ nevertheless corresponds to relatively slow swimming from a biological perspective: According to the approximate calculation in~\cite{Morozov:PRX:2020}, $L$ for \emph{E. coli} bacteria lies somewhere in the range between 5 and 20, indicating that the effect of Brownian motion on active diffusion is likely negligible in suspensions of swimming bacteria due to their fast self-propulsion. In our LB simulations, studying values higher than $L \approx 4$ is challenging, as these large swimming speeds both require very large systems to avoid significant finite-size effects and yields artifacts due to the effect of finite Reynolds number in the swimmer-tracer scattering dynamics~\citep{deGraaf1}. We nevertheless  numerically studied $\xi (\mathrm{Pe})$ using Eq.~\eqref{eq:DA_exact} for larger values of $L$, verifying that both the peak and the subsequent decrease in $\xi$ continues to move to even lower values of Pe as $L$ is increased.

\begin{figure*}
\centering
\begin{minipage}[t]{0.47\textwidth}
    \includegraphics[width=\textwidth, clip, trim=0.0cm 0.0cm 0.0cm 0.0cm]{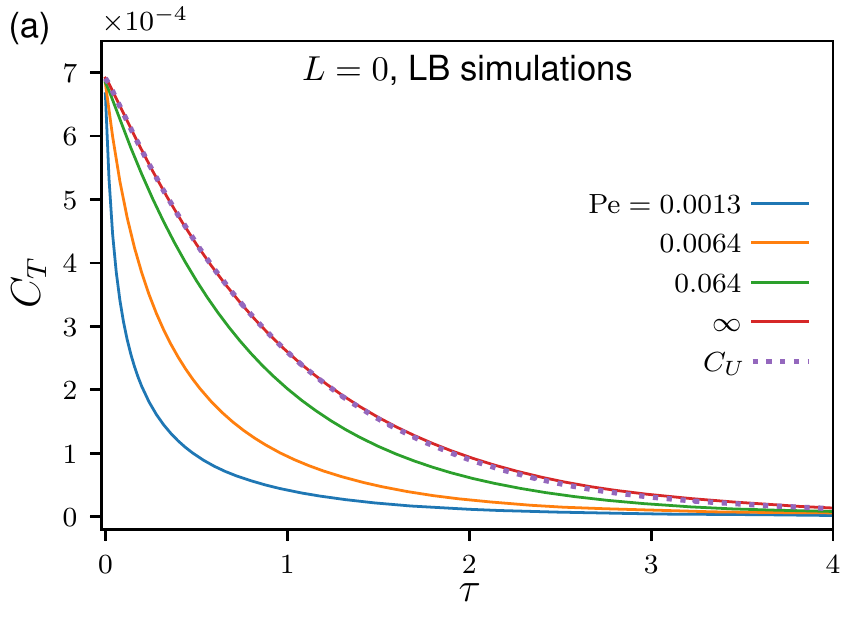}
    \includegraphics[width=\textwidth, clip, trim=0.0cm 0.0cm 0.0cm 0.0cm]{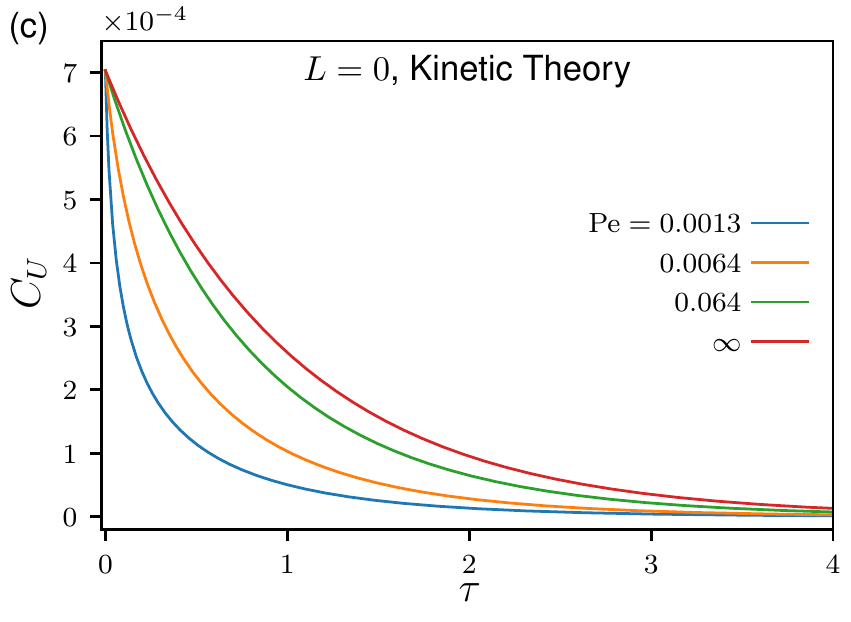}
\end{minipage}
\hfill
\begin{minipage}[t]{0.47\textwidth}
    \includegraphics[width=\textwidth, clip, trim=0.0cm 0.0cm 0.0cm 0.0cm]{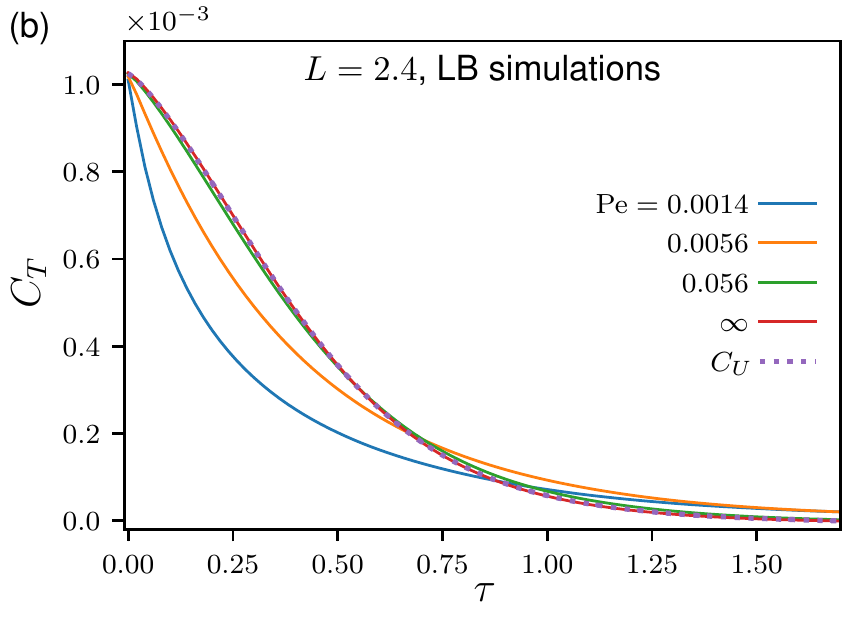}
    \includegraphics[width=\textwidth, clip, trim=0.0cm 0.0cm 0.0cm 0.0cm]{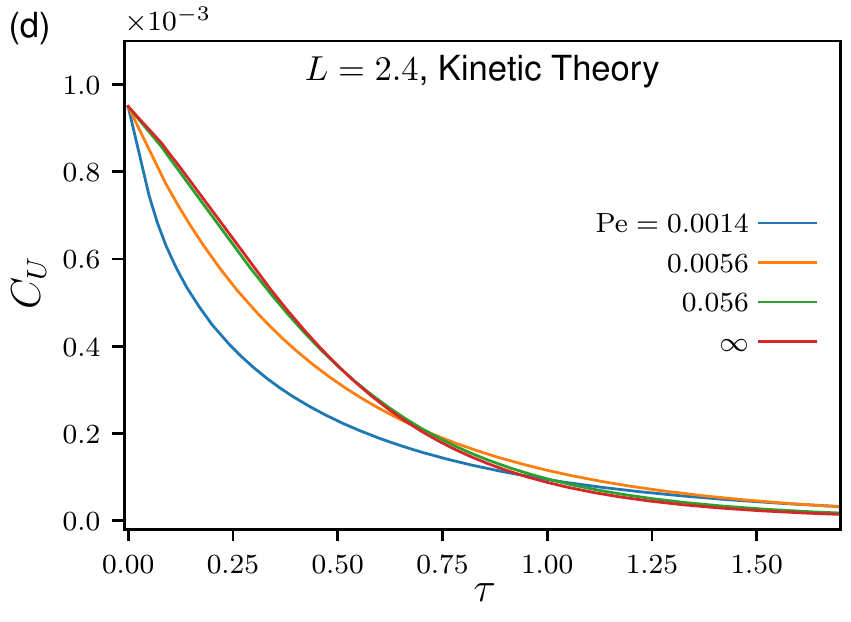}
\end{minipage}
\caption{\textbf{Brownian motion decorrelates tracer trajectories.} Panels (a) and (b) show the fluid velocity autocorrelation $C_T(t)$ in the co-moving tracer frame measured from LB simulations for (a) shakers ($L=0$) and (b) swimmers with $L = 2.4$ at indicated values of Pe. The dotted line shows the correlation function $C_U(t)$ of the fluid velocity in the lab frame, demonstrating that the stationary-tracer approximation $C_T(t) \approx C_U(t)$ is excellent in the absence of Brownian tracer diffusion (Pe $\rightarrow \infty$). Panels (c) and (d) show the corresponding lab-frame correlation function $C_U(t)$, obtained from kinetic theory (Eq.~\eqref{eq:DA_exact}) for a suspension of diffusing swimmers, as described in Section~\ref{sec:Theory}. $C_T$ and $C_U$ are nondimensionalised using $\kappa$ and $\varepsilon$. }

\label{fig:U_correlation}
\end{figure*}

To further understand the mechanism behind the reduction in $D_A$ with $D_0$, we consider the two autocorrelation functions $C_T(t)$ and $C_U(t)$, which respectively measure the fluid autocorrelation in the co-moving tracer frame and in the lab frame. Figure~\ref{fig:U_correlation} shows these correlation functions for $L = 0$ and  $L = 2.4$, with the top row corresponding to LB results for $C_T$ and the bottom row to kinetic theory results from Eq.~\eqref{eq:CU} for $C_U$. First, we notice that the two sets of curves are very similar, implying that the stationary tracer approximation $C_T \approx C_U$ is indeed accurate. Secondly, we notice that the decay of the correlation function is significantly faster for swimmers than for shakers, again illustrating that self-propulsion acts an efficient decorrelation mechanism for $\UU$. The effect of finite $D_0$ for shakers (left column) is simply to decrease the relaxation time of the exponential decay, in accordance with the $L = 0$ limit of Eq.~\eqref{eq:CU}. For swimmers, the situation is more complex: For short times, the flow field decays faster with decreasing Pe, while the long-time tail of $C_T$ and $C_U$ instead becomes somewhat more extended with decreasing Pe. For fast enough swimmers, the latter effect leads to the local maximum at $\xi > 1$ for intermediate Pe observed in Fig.~\ref{fig:xi_Pe}b for the two highest $L$. Finally, we note that the equal-time fluid velocity variance $\langle U^2(\rr_T) \rangle$, corresponding to the $t = 0$ values of $C_T$ and $C_U$, is independent of Pe. This means that, regardless of the ratio between diffusive and active motion, the tracer particles sample the overall flow field homogeneously. This fact is non-trivial, since dry active particles that move autonomously on a solid substrate are known to preferentially sample regions where they move slowly~\citep{Stenhammar:SciAdv:2016}. Our results thus highlight that this generic mechanism is absent for inertialess point tracers advected by an incompressible fluid. However, for microswimmer systems where entrainment due to tracers being captured by the near-field flows of passing swimmers~\citep{Polin:NatComm:2016}, we would expect the fluid flow sampled by tracers to be significantly different from the average flow field in the system.

\begin{figure*} 
\centering
\begin{minipage}[t]{0.49\textwidth}
\includegraphics[width=\textwidth, clip, trim=0.1cm 0.1cm 0.1cm 0.1cm]{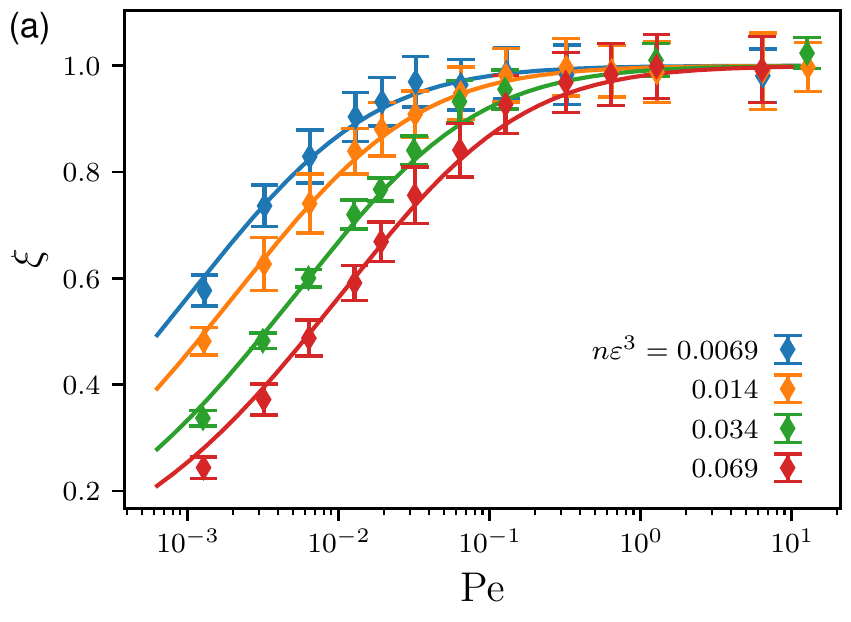}
\end{minipage}
\hfill 
\begin{minipage}[t]{0.49\textwidth}
\includegraphics[width=\textwidth, clip, trim=0.1cm 0.1cm 0.1cm 0.1cm]{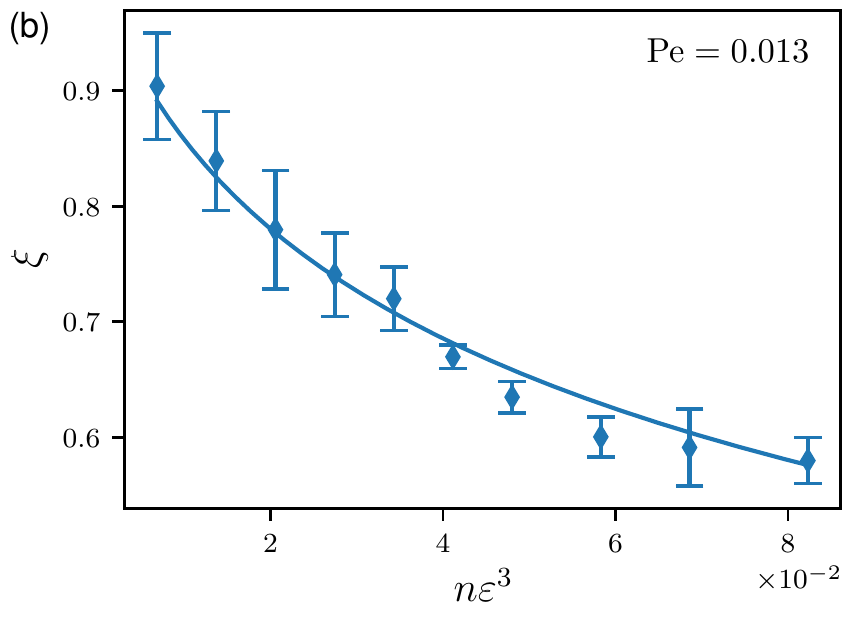}
\end{minipage}
\caption{\textbf{Reduction of $D_A$ varies with microswimmer density.} Panel (a) shows $\xi$ as a function of Pe for shakers ($L=0$) at various densities $n$, as indicated, while panel (b) shows $\xi$ as a function of $n$, at a fixed value of $\mathrm{Pe} = 0.013$. Simulation data are given by the symbols, with error bars obtained from averaging over four separate runs with different initial conditions, while solid lines are computed from Eq. \eqref{eq:DA_exact}.} 

\label{fig:xi_n}
\end{figure*}

In Fig.~\ref{fig:xi_n}, we study the dependence of the suppression of active diffusion on the microswimmer density $n$ in the shaker limit $L=0$. At first sight, this dependence might appear trivial, since $D_A$ is well-known to be linearly dependent on $n$~\citep{Jepson1,Mino1} in the limit of non-interacting swimmers, a fact which is unaffected by Brownian motion as shown by Eq.~\eqref{eq:DA_exact}. Since $\xi$ measures the ratio between the Brownian and non-Brownian values of $D_A$, one would na\"ively expect $\xi$ to be independent of $n$. However, since the P\'eclet number itself, as defined in Eq.~\eqref{eq:Pe_Def}, increases with $n$ for constant $D_0$, the relative effect of Brownian motion on $D_A$ is in fact a complex function of $n$ even for noninteracting swimmers, as shown in Fig.~\ref{fig:xi_n}. More specifically, according to Fig.~\ref{fig:xi_n}b the suppression of active diffusion becomes more significant with increasing microswimmer density. In physical units, the highest concentration considered ($n\varepsilon^3 \approx 0.08$) approximately corresponds to a bacterial concentration of $10^9$ mL$^{-1}$, which is somewhat higher than the highest \emph{E. coli} concentration considered by \cite{Jepson1} but still within the range of concentrations where swimmer-swimmer correlations are reasonably small~\citep{Stenhammar1}. In summary, our results thus show that, for Brownian diffusion to have any measurable influence on the hydrodynamic diffusion, it is necessary to create a system with a relatively high density of very slow microswimmers; as we discuss in Section~\ref{sec:Conclusions}, this set of parameters is likely not achievable for suspensions of biological microswimmers. 

\section{Summary and conclusions}\label{sec:Conclusions}

In this study we have demonstrated a number of theoretical and computational results regarding the effect of Brownian diffusion on the swimmer-induced hydrodynamic diffusion of tracer particles in a suspension of dipolar microswimmers. Our key finding is that the effect of Brownian diffusivity $D_0$ on the activity-induced, hydrodynamic diffusivity $D_A$ is only significant when the P\'eclet number as defined by Eq.~\eqref{eq:Pe_Def} is below unity, meaning that Brownian diffusivity needs to dominate over the hydrodynamic one. However, the necessary requirement $\mathrm{Pe} < 1$ is only sufficient in the shaker limit $v_s \rightarrow 0$: for swimmers with persistence lenghts larger than the organism size, significantly lower values of Pe are required to perturb the tracer trajectories sufficiently to affect $D_A$. This conclusion is analogous to the independence of $D_A$ on the tumbling rate $\lambda$ for large $v_s$ illustrated in Eq.~\eqref{eq:DA_approx}: Whenever $v_s$ is large, the decorrelation of $\UU$ by swimming will dominate over the decorrelation due to tumbling and translational diffusion, and the dependence on $\lambda$ and $D_0$ will thus vanish in the limit $v_s \rightarrow \infty$. While this effect is expected, what is perhaps surprising is the rather moderate values of $L = v_s / (\varepsilon \lambda)$ necessary to render the coupling between $D_0$ and $D_A$ negligible, as illustrated in Fig.~\ref{fig:xi_Pe}b. To put these values into perspective, we use the conservative estimate $L = 5$ for \emph{E. coli}. By virtue of Fig.~\ref{fig:xi_Pe}b, we require that $\mathrm{Pe} \leq 10^{-3}$ for Brownian motion to have an effect of $\sim 5$ percent on $D_A$. Using as an example the minimum value $D_A \approx \SI{d-2}{\square\micro\meter\per\second}$ measured by \cite{Jepson1} in a 3-dimensional \emph{E. coli} suspension, this requirement thus implies that $D_0 \geq \SI{10}{\square\micro\meter\per\second}$, which by virtue of the Stokes-Einstein relation corresponds to a tracer radius of $R_0 \sim 20$ nm. While this is significantly smaller than used in typical measurements on colloidal tracers~\citep{Mino1,Patteson1,Leptos1}, this diffusion coefficient is close to the value of $D_0$ measured for dextran in \emph{E. coli} suspensions by~\cite{Kim1}. It is also fully feasible to realise such low P\'eclet numbers for micron-sized spheres by instead decreasing the bacterial density to very low values; however, measuring the correction to $D_A$ for $\mathrm{Pe} \sim 10^{-3}$ represents a major difficulty, since it amounts to measuring a $\sim$ 5 percent deviation of an effective diffusivity which is itself a thousand times smaller than the Brownian diffusion. It would thus require an extremely accurate determination of $D_0$, which then needs to be subtracted from the total measured diffusion constant to determine $D_A$. Obtaining this accuracy in a colloidal suspension would be very challenging due to particle polydispersity, interactions with boundaries, temperature gradients, and other system-specific complications. We thus conclude that, for typical $L$ values relevant for biological microswimmers, our results imply that the effect of Brownian motion on $D_A$ is likely negligible for all practical purposes. To experimentally observe the reduction in $D_A$, one would instead need to study a system of dipolar shakers, which stir up the surrounding fluid without self-propelling. While this is a somewhat exotic type of system, it could potentially be realised by anchoring molecular motors or biological microswimmers to a surface. In a biological setting, the shaker limit furthermore  resembles previously developed models of enzymes anchored to lipid bilayers that induce dipolar flows through cyclical conformation changes~\citep{Hosaka:SoftMatter:2020}.

Our results are qualitatively consistent with the previous theoretical results obtained by~\cite{Kasyap2014}, including the non-monotonic behaviour of $D_A$ with respect to $D_0$ at high swimming speed. Their results however differ in three important ways. First of all, their results consider a ``slender swimmer'' model, where the fluid is forced via a stress applied along a line representing the swimmer body, rather than by two point forces. While this model should lead to a dipolar flow in the far field, the near-field differences are significant, as illustrated in Fig. 8 of~\cite{Kasyap2014}. Furthermore, they consider only the fast-swimming limit where, according to Eq.~\eqref{eq:DA_approx}, $D_A$ for dipolar microswimmers is independent of $\lambda$. Finally, they parametrise their model in a qualitatively different way than us: as discussed above, they adopt a definition of $\mathrm{Pe}$ based on $v_s$ rather than on the tracer diffusivity as in Eq.~\eqref{eq:Pe_Def}. In this description, $v_s$ is furthermore directly coupled to $\kappa$, so that changing Pe simultaneously changes the activity of the bath (via $\kappa$) and the swimming speed $v_s$. These differences makes it difficult to compare directly with our results, as we consider the effects of fluid advection and self-propulsion separately via Pe and $L$. 

Thus, even though Brownian motion is unlikely to provide a significant dependence of $D_A$ on $R_0$ for tracers immersed in suspensions of biological microswimmers, there are several other mechanisms that need to be studied to explain the non-monotonic dependence observed experimentally~\citep{Patteson1} and computationally~\citep{Dyer:PoF:2021}. First of all, the effect of tracer entrainment by the near field flows of the swimmer is strongly dependent on the size ratio between the swimmer and the tracer~\citep{Polin:NatComm:2016}, although we expect this term to be small for micron-sized tracer particles in \emph{E. coli} suspensions. Secondly, the finite size of the tracer will change the equation of motion~\eqref{eq:rdot_tr} into the Fax\'en equation that takes into account the non-linearity of the flow field~\citep{Kim-Karrila}, an effect which was implicitly included in the wavelet Monte Carlo simulations by~\cite{Dyer:PoF:2021} and, together with tracer entrainment, is a significant explanation of their observed $R_0$ dependence of $D_A$. Finally, non-hydrodynamic interactions such as direct collisions, electrostatic interactions, and artifacts due to container walls are likely to depend in a non-trivial manner on the tracer size for each system in question. We thus conclude by noting that significant further experimental and theoretical work is necessary to disentangle the system-specific properties from the generic properties of tracer dynamics in microswimmer suspensions. 

\section*{Acknowledgement}
Discussions with Cesare Nardini are kindly acknowledged. The computations were enabled by resources provided by LUNARC. For the purpose of open access, the authors have applied a Creative Commons Attribution (CC BY) licence to any Author Accepted Manuscript version arising from this submission. 

\section*{Funding} 
This work was financed through the Knut and Alice Wallenberg Foundation (project grant KAW 2014.0052). JS acknowledges financial support from the Swedish Research Council (Project No. 2019-03718). 

\section*{Competing Interests}
The authors report no competing interests.

\appendix
\section{Kinetic theory for $D_A(D_0)$. } 
\label{App:CT}

In this Appendix, we demonstrate the main steps in the derivation of Eq.~\eqref{eq:CU}. The analysis follows closely a similar derivation presented in~\cite{Morozov:PRX:2020}, albeit with two major differences. First, in addition to that work, we include the effect of microswimmer Brownian diffusivity, as discussed in Section \ref{sec:Theory}. Second, we consider the case of non-interacting microswimmers, which significantly simplifies the analysis. Due to the similarity with the derivation in~\cite{Morozov:PRX:2020}, we here present the key steps of the derivation and refer the interested reader to that paper for technical details.

The quantity of interest is the fluid velocity autocorrelation function $C_U(t)$, formally defined as
\begin{align}
C_U(t) = \lim_{t'\rightarrow\infty} \frac{1}{V}\int d{\bm r} \,\overline{U^{\alpha}\left( {\bm r}, t' \right) U^{\alpha}\left( {\bm r}, t+t' \right)},
\end{align} 
where $V$ is the volume of the system, and the overbar denotes an average over the stochastic history, \emph{i.e.}, the history of tumble events, and the long-time limit ensures independence of the initial state of the system. Here and in the following, the superscript indices denote Cartesian components of vectors. For a given state of the system, the instantaneous fluid velocity $U^{\alpha}$ at a position $\rr$ and time $t$ is readily obtained as a superposition of individual velocity fields generated by the microswimmers
\begin{align}
&U^{\alpha}\left( \rr, t \right) = \sum_{i=1}^N u_s^\alpha({\rr}; {\rr}_i(t),{\pp}_i(t)).
\end{align}
Here, $\rr_i(t)$ gives the instantaneous position of particle $i$, while the unit vector $\pp_i(t)$ gives its instantaneous orientation; the index $i= 1 \dots N$ enumerates the particles, where $N$ is the total number of microswimmers. In the following, we assume $\uu_s$ to be given by the regularised hydrodynamic dipole \eqref{eq:u_s}.

Time evolution of the suspension comprises spatial motion of the microswimmers according to Eq.~\eqref{eq:eom} and their random re-orientation with rate $\lambda$, as discussed in Section~\ref{section:introduction}. These dynamics can be alternatively described by the master equation
\begin{align}
\partial_t F_N  + \sum_{i=1}^N \Big( v_s p_i^\alpha \nabla_i^\alpha - D_0  \nabla_i^2 \Big) F_N
 = - N \lambda F_N +  \frac{\lambda}{4\pi} \sum_{i=1}^N \int \mathrm{d} \pp_i F_N,
\label{Liouville}
\end{align}
where $F_N=F_N({\rr}_1,\dots,{\rr}_N,{\pp}_1,\dots,{\pp}_N,t)$ is the $N$-particle probability distribution function. Here, ${\bm \nabla}_i$ denotes spatial derivatives with respect to the coordinates of particle $i$. 
As shown in~\cite{Morozov:PRX:2020}, the same dynamics can be conveniently encoded by the following equation:
\begin{align}
\partial_t h + v_s p^\alpha \nabla^\alpha h - D_0  \nabla^2 h + \lambda h
- \frac{\lambda}{4\pi} \int d\pp \, h = \chi(\rr,\pp, t),
\label{Eqh}
\end{align}
where $h=h(\rr,\pp, t)$ is an auxiliary field related to the Klimontovich correlation function~\citep{Silin1962}, and $\chi$ is a noise term with the following properties:
\begin{align}
& \langle \chi(\rr,\pp,t) \rangle = 0, 
\label{noiseaverage}\\
& \langle \chi(\rr,\pp, t) \chi(\rr',\pp', t')\rangle = \frac{n}{4\pi} \delta(t-t') \Bigg[ 2 \lambda   \delta(\rr - \rr') 
\left( \delta(\pp - \pp' ) - \frac{1}{4\pi} \right) \nonumber \\
&\qquad\qquad\qquad\qquad\qquad\qquad\qquad\qquad\qquad 
- D_0 \delta(\pp - \pp' ) \Big( \nabla^2 + \nabla'^{2}\Big)    \delta(\rr - \rr' )  \Bigg].
\label{noisevariance}
\end{align}
The advantage of this representation lies in its direct relation to the phase-space density correlation function~\citep{Silin1962}, which allows us to express the fluid velocity autocorrelation function as
\begin{align}
&C_U(t) = \lim_{t'\rightarrow\infty} \frac{1}{V}\int d\rr \int d\rr'd\rr''  d\pp'd\pp''  
u_s^\alpha(\rr; \rr',\pp') u_s^\alpha(\rr; \rr'',\pp'')  \nonumber \\
& \qquad \qquad\qquad \qquad \qquad \qquad
 \times \langle h(\rr',\pp', t) h(\rr'',\pp'', t+t') \rangle_\chi,
\label{CRTtmp}
\end{align} 
where the angular brackets denote the average of the (fictitious) noise $\chi$ (see \cite{Morozov:PRX:2020} for details).
~
The linear equation \eqref{Eqh} is readily solved by introducing the Fourier--Laplace transform $\hat{h}$ of the auxiliary field $h$:
\begin{align}
\hat{h}({\bm k}, {\bm p}, s) = \int_0^\infty dt e^{-s t}  \int d{\bm r} e^{i {\bm k}\cdot{\bm r}} h({\bm r}, {\bm p}, t),
\end{align}
which yields
\begin{align} \label{eq:hsol}
\hat{h}({\bm k}, {\bm p}, s) = \frac{\hat{\chi}({\bm k}, {\bm p}, s)}{\sigma({\bm k}, {\bm p}, s)} + \frac{\lambda}{4\pi \sigma({\bm k}, {\bm p}, s)}
\frac{\int d{\bm p}' \frac{\hat{\chi}({\bm k}, {\bm p}', s)}{\sigma({\bm k}, {\bm p}', s)}}{1-\frac{\lambda}{4\pi} \int \frac{d{\bm p}' }{\sigma({\bm k}, {\bm p}', s)}}.
\end{align}
Here,  $\sigma({\bm k},{\bm p}, s) = s + \lambda + D_0 k^2 + i v_s ({\bm k}\cdot{\bm p})$, $\hat{\chi}({\bm k}, {\bm p},s)$ is the Fourier--Laplace transform of the noise, and 
we have dropped the initial condition $\hat{h}({\bm k}, {\bm p}, t=0)$ which does not contribute in the large-$t$ limit.

As demonstrated by \cite{Morozov:PRX:2020}, only the first term in Eq.~\eqref{eq:hsol} contributes to $C_U(t)$. Performing the Fourier--Laplace transform in Eq.~\eqref{CRTtmp} and combining it with Eq.~\eqref{eq:hsol} yields
\begin{align}
&C_U(t) = \frac{ n \kappa^2}{16\pi^4 } \lim_{t'\rightarrow\infty} {\mathcal L}^{-1}_{s_1,t'} {\mathcal L}^{-1}_{s_2,t'+t} 
 \int d{\bm k}  \frac{A^2(k\epsilon)}{k^4} \left( \lambda + D_0 k^2\right)\nonumber \\
&\quad \times \int d{\bm p} \,
(\bm k \cdot {\bm p})^2 \left[1- \frac{(\bm k \cdot {\bm p})^2}{k^2}\right] \frac{1}{s_1+s_2}\nonumber \\
&  \qquad\qquad \times  
\frac{1}{\lambda + s_1 + D_0 k^2 + i v_s (\bm k\cdot {\bm p})} 
\frac{1}{\lambda + s_2 + D_0 k^2 - i v_s (\bm k\cdot {\bm p})}.
\label{C0Laplace}
\end{align}
Performing the inverse Laplace transforms, denoted symbolically by ${\mathcal L}^{-1}$, and integrating over $\bm p$ and the orientation of $\bm k$, we finally arrive at Eq.~\eqref{eq:CU} in the main text.

\bibliographystyle{jfm}

\bibliography{refs_anisotropy} 

\end{document}